\documentstyle[12pt]{article}
\vspace{1.8cm} \oddsidemargin 0pt \evensidemargin 0pt  \topmargin
-1.5cm \footheight 26pt \footskip 10mm

\begin{document}
\baselineskip 20pt
\begin{center}
\baselineskip=24pt {\Large\bf The  optical phenomena in
multiple-photon interactions (I) }

\vspace{1cm} \centerline{Xiangyao Wu$^{a}$
\footnote{E-mail:wuxy2066@163.com}}Pi-Feng Gong$^{a}$,
 Zhi-Yong Yi $^{b}$, Chang-Rui Wu$^{a}$, Xiao-Yan Gu$^{c}$ and Zong-Hua Shi$^{a}$

\vspace{0.8cm}

\noindent{\footnotesize a. \textit{Institute of Physics ,
Qufu Normal University, Qufu 273165, China}}\\
{\footnotesize b. \textit{BeiJing Institute of PetroChemical TECHN
,P.O.Box 8001, BeiJing 102617, China}}\\
{\footnotesize c. \textit{Institute of High Energy Physics,
P.O.Box 918(4), Beijing 100039, China}}
\end{center}
\date{}
\renewcommand{\thesection}{Sec. \Roman{section}} \topmargin 10pt
\renewcommand{\thesubsection}{ \arabic{subsection}} \topmargin 10pt
{\vskip 5mm
\begin {minipage}{140mm}
\centerline {\bf Abstract} \vskip 8pt
\par

\indent\\

In the recent experiment, the phenomena of superluminal and
slow-light propagation in dispersive medium were found, and there
are various explanation in theory. We find the phenomenon can be
explained by multiple-photon interaction. Otherwise, the
multiple-photon interaction can also appear other optical
phenomenon: doubling frequency , sum frequency, difference
frequency which are found in nonlinear optics.
\end {minipage}

\vspace*{2cm} {\bf PACS number(s): 63.20.Kr, 14.70.Bh, 42.65.-k,
42.50.Gy }
\newpage

{\bf 1. Introduction} \vskip 8pt

Since laser appeared in 1960's, many new optical phenomena were
found in nonlinear optics, e.g. frequency doubling, sum frequency,
many-photo absorption, light condense, ultra-short impulse, laser
self-interaction and so on \cite{s1}. These phenomena are
generated when strong light interact with the nonlinear medium. In
the past years, the new optical phenomena of superluminal and
slow-light were found in the dispersion medium
\cite{s2}\cite{s3}\cite{s4}\cite{s5}\cite{s6}, and there are
various explanation in theory. In this paper, we research these
new optical phenomena with multiple-photon interaction,
photon-photon interaction belongs to electromagnetic interaction,
and it must satisfy some conservation laws, one of them is $C$
parity conservation. For photon, $C$ parity $\eta_{c}=-1$. So,
 photon-photon interaction should satisfy \cite{s7}: \\
(1) if the results of photon-photon interaction  generate photons,
then even photons  will generate even photons and odd photons
 will generate odd photons.\\
(2) if the results of photon-photon generate material particle
 ($m_{0}\neq 0$), then they will be generate neutral particle
$n_{0}$ or neutral particle system ($n+\overline{n}$).\\

In quantum field theory, there isn't direct-acting among photons,
and they interact indirectly by Dirac vacuum. More recently, the
noncommutative QED theory has been developed \cite{s10}\cite{s11}.
In the theory, photon has self-interaction unlike the case of
ordinary QED, and Many new predictions \cite{s10}\cite{s11} can be
obtained from the noncommutative QED theory: multiple-photon
direct-acting, superluminal photon, the photons of bound states
and so on. In this paper, we can also obtain many new optical
phenomena from multiple-photon interaction, and they are discussed
only in kinematics. The research from dynamics will be appeared in
the next paper.

{\bf 2. Doubling frequency , sum frequency and difference
frequency} \vskip 8pt In the following, we can obtain the optical
phenomena of doubling
 frequency, sum frequency and difference frequency from
 three-photon collision. We consider two kinds of collision of three-photon.\\
case (a):\\
 \setlength{\unitlength}{0.06in}
  \begin{picture}(100,15)
  \put(26,2){\vector(1,0){8}}
 \put(26,6){\vector(1,0){8}}
 \put(44,4){\vector(-1,0){8}}
 \put(47,3){\\= }
 \put(51,2){\vector(1,0){8}}
 \put(51,6){\vector(1,0){8}}
 \put(68,4){\vector(-1,0){8}}
 \put(29,3){\makebox(2,1)[c]{$h\nu_{2}$}}
 \put(29,7){\makebox(2,1)[c]{$h\nu_{1}$}}
 \put(39,5){\makebox(2,1)[c]{$h\nu_{3}$}}
 \put(54,3){\makebox(2,1)[c]{$h\nu_{5}$}}
 \put(54,7){\makebox(2,1)[c]{$h\nu_{4}$}}
 \put(63,5){\makebox(2,1)[c]{$h\nu_{6}$}}
\end{picture}\\
where $\nu_{i}$ ($i=1,2,\cdots$) is frequency of $i-th$ photon,
the arrows represent moving direction of photon. In the  photon
collision, the energy and momentum are conservative.
\begin{equation}
h(\nu_{1}+\nu_{2}+\nu_{3})=h(\nu_{4}+\nu_{5}+\nu_{6})
\end{equation}
\begin{equation}
h(\frac{1}{\lambda_{1}}+\frac{1}{\lambda_{2}}-\frac{1}{\lambda_{3}})=
h(\frac{1}{\lambda_{4}}+\frac{1}{\lambda_{5}}-\frac{1}{\lambda_{6}})
\end{equation}
 Combining the Eq.(1) and (2)
\begin{equation}
\nu_{3}=\nu_{6}
\end{equation}
\begin{equation}
\nu_{1}+\nu_{2}=\nu_{4}+\nu_{5}
\end{equation}
From Eq.(4) if $\nu_{4}=2\nu_{1}$ then $\nu_{5}=\nu_{2}-\nu_{1}$
and if $\nu_{4}=3\nu_{1}$ then $\nu_{5}=\nu_{2}-2\nu_{1}$. So
three-photo collision may produce frequency doubling, triplex
frequency and difference frequency light phenomena.\\
case (b):\\

 \setlength{\unitlength}{0.06in}
  \begin{picture}(60,15)
  \put(29,2){\vector(1,0){8}}
 \put(29,6){\vector(1,0){8}}
 \put(47,4){\vector(-1,0){8}}
 \put(50,3){\\= }
 \put(62,2){\vector(-1,0){8}}
 \put(62,6){\vector(-1,0){8}}
 \put(64,4){\vector(1,0){8}}
 \put(32,3){\makebox(2,1)[c]{$h\nu_{2}$}}
 \put(32,7){\makebox(2,1)[c]{$h\nu_{1}$}}
 \put(42,5){\makebox(2,1)[c]{$h\nu_{3}$}}
 \put(57,3){\makebox(2,1)[c]{$h\nu_{5}$}}
 \put(57,7){\makebox(2,1)[c]{$h\nu_{4}$}}
 \put(66,5){\makebox(2,1)[c]{$h\nu_{6}$}}
\end{picture}\\
From the conservation law of energy and momentum
\begin{equation}
h(\nu_{1}+\nu_{2}+\nu_{3})=h(\nu_{4}+\nu_{5}+\nu_{6})
\end{equation}
\begin{equation}
h(\frac{1}{\lambda_{1}}+\frac{1}{\lambda_{2}}-\frac{1}{\lambda_{3}})
=h(\frac{1}{\lambda_{6}}-\frac{1}{\lambda_{4}}-\frac{1}{\lambda_{5}})
\end{equation}
From Eq.(5) and Eq.(6)
\begin{equation}
\nu_{6}=\nu_{1}+\nu_{2}
\end{equation}
So three-photon collision can also produce sum frequency light. In
experiment, frequency doubling, many-times frequency, sum
frequency and difference frequency light are usually found in
nonlinear medium \cite{s1}. We think these optical phenomena can
be produced from collision three-photon or more photons.

{\bf 3. Phenomena of superluminal} \vskip 8pt
 The superluminal phenomena  have been found in many
 experiments \cite{s2}. We think the phenomena can be observed in
  three-photon collision, as follows:\\
  case (a):\\

 \setlength{\unitlength}{0.06in}
  \begin{picture}(60,15)
  \put(29,2){\vector(1,0){8}}
 \put(29,6){\vector(1,0){8}}
 \put(47,4){\vector(-1,0){8}}
 \put(49,3){\\= }

 \put(54,4){\vector(1,0){8}}
 \put(32,3){\makebox(2,1)[c]{$h\nu_{2}$}}
 \put(32,7){\makebox(2,1)[c]{$h\nu_{1}$}}
 \put(42,5){\makebox(2,1)[c]{$h\nu_{3}$}}
 \put(57,5){\makebox(2,1)[c]{$h\nu_{4}$}}

\end{picture}\\
From the conservative laws of energy and momentum
\begin{equation}
E=h(\nu_{1}+\nu_{2}+\nu_{3})
\end{equation}
\begin{equation}
P=\frac{h}{\lambda_{1}}+\frac{h}{\lambda_{2}}-\frac{h}{\lambda_{3}}
\end{equation}
Where $E$ and $P$ are energy and momentum of photon $\nu_{4}$\\
The speed of photon $\nu_{4}$ is $\tilde{c}$:
$$\tilde{c}=\frac{E}{P}=
\frac{h(\nu_{1}+\nu_{2}+\nu_{3})}
{h(\frac{1}{\lambda_{1}}+\frac{1}{\lambda_{2}}-\frac{1}{\lambda_{3}})}
=\frac{c(\nu_{1}+\nu_{2}+\nu_{3})}{\nu_{1}+\nu_{2}-\nu_{3}}>c$$\\
 case (b): \\

\setlength{\unitlength}{0.06in}
  \begin{picture}(100,15)
  \put(30,8.5){\vector(1,0){8}}
 \put(39,17){\vector(0,-1){8}}
 \put(39,0){\vector(0,1){8}}
 \put(45,8){\\= }

 \put(49,8.5){\vector(1,0){8}}
 \put(41,12){\makebox(2,1)[c]{$h\nu_{2}$}}
 \put(32,9){\makebox(2,1)[c]{$h\nu_{1}$}}
 \put(41,5){\makebox(2,1)[c]{$h\nu_{3}$}}
 \put(52,9.5){\makebox(2,1)[c]{$h\nu_{4}$}}

\end{picture}\\
Let $\nu_{2}=\nu_{3}$, the energy and momentum of photon $\nu_{4}$
is

\begin{equation}
E=h\nu_{1}+h\nu_{2}+h\nu_{3}
\end{equation}
\begin{equation}
P=\frac{h}{\lambda_{1}}
\end{equation}
 The speed of photon $\nu_{4}$ is
$$\tilde{c}=\frac{E}{P}=\frac{c(\nu_{1}+\nu_{2}+\nu_{3})}{\nu_{1}}>c$$
Obviously,  many-photon ($n>3$) collision can also produce
superluminal photon.

{\bf 4. Phenomena of slow-light}\vskip 8pt More recently, the
phenomena of slow-light were found in many
experiments\cite{s6}\cite{s7}\cite{s8}\cite{s9}. We think it can
also be observed in many-photon collision.
\\
\setlength{\unitlength}{0.06in}
  \begin{picture}(100,15)
  \put(30,5){\vector(1,0){8}}
 \put(46,2.5){\vector(-1,0){8}}
 \put(30,0){\vector(1,0){8}}
 \put(48,2){\\= }
 \put(64,2){\\+ }
 \put(67,2){\\$n_{0}$}

 \put(53,2.7){\vector(1,0){8}}
 \put(40,3.1){\makebox(2,1)[c]{$h\nu_{3}$}}
 \put(32,5.5){\makebox(2,1)[c]{$h\nu_{1}$}}
 \put(32,0.5){\makebox(2,1)[c]{$h\nu_{2}$}}
 \put(55,3.4){\makebox(2,1)[c]{$h\nu_{4}$}}

\end{picture}\\
The photon $\gamma_{4}$ energy and momentum are
\begin{equation}
E=h\nu_{1}+h\nu_{2}+h\nu_{3}-mc^{2}
\end{equation}
\begin{equation}
P=\frac{h}{\lambda_{1}}+\frac{h}{\lambda_{2}}-\frac{h}{\lambda_{3}}\pm
P_{0}
\end{equation}
where $mc^{2}$ and $P_{0}$ are energy and momentum of neutral
particle $n_{0}$. In Eq.(13), $"\pm"$ is from the same or opposite
momentum direction of photon $\nu_{4}$ and particle $n_{0}$. \\
The speed of photon $\gamma_{4}$ is
\\
\begin{equation}
\tilde{c}=\frac{E}{P}=c\frac{\nu_{1}+\nu_{2}+\nu_{3}-\frac{mc^{2}}{h}}
{\nu_{1}+\nu_{2}-\nu_{3}\pm\frac{P_{0}c}{h}}
\end{equation}
if\\
\begin{equation}
\frac{\nu_{1}+\nu_{2}+\nu_{3}-\frac{mc^{2}}{h}}
{\nu_{1}+\nu_{2}-\nu_{3}\pm\frac{P_{0}c}{h}}<1
\end{equation}
then\\
\begin{equation}
\tilde{c}<c
\end{equation}
The Eq. (15) is
\begin{equation}2h\nu_{3}<mc^{2}\pm P_{0}c
\end{equation}\\
When the momentum direction of photon $\nu_{4}$ and particle
$n_{0}$ is same, the Eq. (17) is
\begin{equation}
2h\nu_{3}<mc^{2}-P_{0}c^{2}
\end{equation}

On this condition, photon $\gamma_{4}$ is slow-light.\\
When the momentum direction of photon $\nu_{4}$ and $n_{0}$ is
opposite, the Eq. (17) is

\begin{equation}
2h\nu_{3}<mc^{2}+P_{0}c
\end{equation}
On this condition, photon $\gamma_{4}$ is slow-light. \\

We think that the phenomena of slow-light can also been found in
multiple-photon interaction, which means  there
are slow-light photons in the vacuum.\\

{\bf 5. Particle mass problem} \vskip 8pt
In particle physics,
particle mass is from spontaneous  symmetry breaking and Higgs
mechanism and particles can be produced by photon-photon
interaction and they can be annihilation into photons. We can give
particle rest mass $m_{0}$ in photon-photon
interaction. We can consider it as follows:\\
case (a): two-photon collision\\
\setlength{\unitlength}{0.06in}
\begin{picture}(100,15)
 \put(38,3){\vector(1,0){8}}
  \put(56,3){\vector(-1,0){8}}

 \put(58,2){\\= }

 \put(61,2){\\$n_{0}$}

 \put(50,3.5){\makebox(2,1)[c]{$h\nu_{2}$}}
 \put(42,3.5){\makebox(2,1)[c]{$h\nu_{1}$}}

\end{picture}\\

$$\nu_{1}+\nu_{2}\longrightarrow n_{0}$$
From the conservative laws of energy and momentum\\
\begin{equation}
E=h\nu_{1}+h\nu_{2}
\end{equation}
\begin{equation}
P=\frac{h}{\lambda_{2}}-\frac{h}{\lambda_{1}}
\end{equation}
Where $E$ and $P$ are energy and momentum of particle $n_{0}$.
From special relatively, the relation between the total energy
$E$ of particle $n_{0}$ and its momentum $P$ and rest mass $m_{0}$ is\\
\begin{equation}
E^{2}=m_{0}^{2}c^{4}+c^{2}P^{2}
\end{equation}
From Eq.(20), (21) and (22) \\
$$E_{0}^{2}=m_{0}^{2}c^{4}=E^{2}-c^{2}P^{2}=
h^{2}(\nu_{1}+\nu_{2})^{2}-c^{2}
(\frac{h(\lambda_{2}-\lambda_{1})}{\lambda_{1}\lambda_{2}})^{2}
=4h^{2}\nu_{1}\nu_{2}$$ \\
and so\\
\begin{equation}
m_{0}=\frac{2h\sqrt{\nu_{1}\nu_{2}}}{c^{2}}=
\frac{2\sqrt{\varepsilon_{1}\varepsilon_{2}}}{c^{2}}
\end{equation}
The kinetic energy $T$ of particle $n_{0}$ is\\
\begin{equation}
T=E-E_{0}=h\nu_{1}+h\nu_{2}-2h\sqrt{\nu_{1}\nu_{2}}
=h(\sqrt{\nu_{1}}-\sqrt{\nu_{2}})^{2}
\end{equation}\\
case (b): Multiple-photon collision\\
\setlength{\unitlength}{0.06in}
  \begin{picture}(100,15)
  \put(38,9){\vector(1,0){8}}
  \put(38,6){\vector(1,0){8}}
  \put(38,0){\vector(1,0){8}}
 \put(56,9){\vector(-1,0){8}}
 \put(56,6){\vector(-1,0){8}}
 \put(56,0){\vector(-1,0){8}}

 \put(58,5){\\= }
 \put(42,3){\\{$\vdots$}}
 \put(52,3){\\{$\vdots$}}

 \put(61,5){\\$n_{0}$}
 \put(42,9.5){\makebox(2,1)[c]{$h\nu_{1}$}}
 \put(42,6.5){\makebox(2,1)[c]{$h\nu_{2}$}}
 \put(42,0.8){\makebox(2,1)[c]{$h\nu_{m}$}}
 \put(52,9.5){\makebox(2,1)[c]{$h{\nu_{1}}^{'}$}}
 \put(52,6.5){\makebox(2,1)[c]{$h{\nu_{2}}^{'}$}}
 \put(52,0.8){\makebox(2,1)[c]{$h{\nu_{n}}^{'}$}}

\end{picture}\\

$$(\nu_{1}+\nu_{2}+\cdots+\nu_{m})+(\nu_{1^{'}}+\nu_{2^{'}}
+\cdots+\nu_{n^{'}})\longrightarrow n_{0}$$ \\
The total
energy $E$ and momentum $P$ of particle $n_{0}$ are:\\
\begin{equation}
E=h\nu_{1}+h\nu_{2}+\cdots+h\nu_{m}+h\nu_{1^{'}}+h\nu_{2^{'}}
+\cdots+h\nu_{n^{'}}
\end{equation}
\begin{equation}
P=\frac{h}{\lambda_{1}}+\frac{h}{\lambda_{2}}+\cdots+\frac{h}{\lambda_{m}}
-\frac{h}{\lambda_{1^{'}}}-\frac{h}{\lambda_{2^{'}}}-\cdots-\frac{h}{\lambda_{n^{'}}}
\end{equation}\\
and
\begin{equation}
E^{2}=m_{0}^{2}c^{4}+c^{2}P^{2}
\end{equation}
From Eq. (25), (26), and (27) \\
\begin{eqnarray}E_{0}^{2}=m_{0}^{2}c^{4}&&=E^{2}-c^{2}P^{2}\nonumber\\&&
=h^{2}(\nu_{1}+\nu_{2}+\cdots+\nu_{m}+\nu_{1^{'}}+\nu_{2^{'}}\nonumber\\&&
+\cdots+\nu_{n^{'}})^{2}-h^{2}(\nu_{1}+\nu_{2}+\cdots+\nu_{m}-\nu_{1^{'}}-\nu_{2^{'}}
-\cdots-\nu_{n^{'}})^{2}\nonumber\\&&
=4h^{2}(\nu_{1}+\nu_{2}+\cdots+\nu_{m})(\nu_{1^{'}}+\nu_{2^{'}}
+\cdots+\nu_{n^{'}})\nonumber
\end{eqnarray}\\
and so
\begin{eqnarray}
m_{0}=\frac{E_{0}}{c^{2}}&&=\frac{2h\sqrt{(\nu_{1}+\nu_{2}+\cdots+\nu_{m})
(\nu_{1^{'}}+\nu_{2^{'}}+\cdots+\nu_{n^{'}})}}{c^{2}}\nonumber\\&&
=\frac{2\sqrt{(\varepsilon_{1}+\varepsilon_{2}+\cdots+\varepsilon_{m})
(\varepsilon_{1^{'}}+\varepsilon_{2^{'}}+\cdots+\varepsilon_{n^{'}})}}{c^{2}}
\end{eqnarray}
where $\varepsilon_{i}=h\nu_{i}$, $\varepsilon_{j}=h\nu_{j^{'}}$
are energy of photons whose frequency are $\nu_{i}$,
$\nu_{j^{'}}$.\\
The kinetic energy of particle $n_{0}$ :
\begin{eqnarray}
T=E-E_{0}&&
=h(\nu_{1}+\nu_{2}+\cdots+\nu_{m}+\nu_{1^{'}}+\nu_{2^{'}}
+\cdots+\nu_{n^{'}})\nonumber\\&&
-2h\sqrt{(\nu_{1}+\nu_{2}+\cdots+\nu_{m})(\nu_{1^{'}}+\nu_{2^{'}}
+\cdots+\nu_{n^{'}})}\nonumber\\&&
=h(\sqrt{\nu_{1}+\nu_{2}+\cdots+\nu_{m}}-\sqrt{\nu_{1^{'}}+\nu_{2^{'}}
+\cdots+\nu_{n^{'}}})^{2}
\end{eqnarray}

So all matter particle ($m_{0}\neq 0$) can be produced in photons
collisions. Its rest mass $m_{0}$  is from the energy of
interaction photons, and its kinetic energy is the rest energy of
photons after those photons generate matter particle.

{\bf 6. Conclusion} \vskip 8pt

   There are a number of new optical phenomena were found in
nonlinear medium. We think these phenomena can be explained by
multiple-photon interaction. Otherwise, matter particle can be
obtained by the energy of interaction photons. These predictions
can be checked in photon collider, but the dynamics mechanism of
multi-photon interaction will be further research.

\newpage
\end{document}